\documentclass[aps,twocolumn,showpacs,superscriptaddress]{revtex4}
\usepackage{epsfig}
\usepackage{dcolumn}
\begin{document}


\title{Giant radio-frequency magnetoabsorption effect in the
cobaltite ceramic La$_{0.5}$Sr$_{0.5}$CoO$_3$}


\author{B. I. Belevtsev}
\email[]{belevtsev@ilt.kharkov.ua}
\affiliation{B. Verkin Institute for Low Temperature Physics and Engineering,
National Academy of Sciences, Kharkov 61103, Ukraine}

\author{A. Ya. Kirichenko}
\affiliation{A. Usikov Institute for Radiophysics and Electronics,
National Academy of Sciences, Kharkov 61085, Ukraine}

\author{N. T. Cherpak}
\email[]{cherpak@ire.kharkov.ua}

\affiliation{A. Usikov Institute for Radiophysics and Electronics,
National Academy of Sciences, Kharkov 61085, Ukraine}

\author{G. V. Golubnichaya}
\affiliation{A. Usikov Institute for Radiophysics and Electronics,
National Academy of Sciences, Kharkov 61085, Ukraine}

\author{I. G. Maximchuk}
\affiliation{A. Usikov Institute for Radiophysics and Electronics,
National Academy of Sciences, Kharkov 61085, Ukraine}

\author{A. B. Beznosov}
\affiliation{B. Verkin Institute for Low Temperature Physics and Engineering,
National Academy of Sciences, Kharkov 61103, Ukraine}

\author{V. B. Krasovitsky}
\affiliation{B. Verkin Institute for Low Temperature Physics and Engineering,
National Academy of Sciences, Kharkov 61103, Ukraine}

\author{P. P. Pal-Val}
\affiliation{B. Verkin Institute for Low Temperature Physics and Engineering,
National Academy of Sciences, Kharkov 61103, Ukraine}

\author{I. N. Chukanova}
\affiliation{Institute for Single Crystals, National Academy of Sciences,
Kharkov 61001, Ukraine}



\begin{abstract}
The DC transport properties of and the radio-frequency (RF) wave
absorption (at 1.33 MHz) in a ceramic sample of
La$_{0.5}$Sr$_{0.5}$CoO$_{3-\delta}$ are measured. The Curie
temperature,  $T_{\mathrm{c}}$, of the sample is about 250~K. A
giant negative magnetoabsorption effect is found. In the vicinity
of $T_{\mathrm{c}}$, the absolute value of the magnetoabsorption
is about 38\% in the  rather low magnetic field 2.1 kOe. This
differs drastically from the measured DC magnetoresistance (MR)
$\delta (H) =[R(0)-R(H)]/R(0)$ which is a mere
0.26 \% near $T_{\mathrm{c}}$ in the same field and
increases to about 2.15~\% in 
$H=20$~kOe. The phenomenon can be understood taking into account
that the magnetoabsorption is determined by influence of magnetic
field on the conductivity and the magnetic permeability, while the
MR is determined solely by the former. The magnetoabsorption
effect can be used to develop  RF devices controlled by
magnetic field and temperature.
\end{abstract}

\pacs{72.80.Ga; 75.30.Vn; 78.70.Gd}

\maketitle

\section{Introduction}
In recent years much attention has been given to mixed-valence
cobaltites of the type La$_{1-x}$Sr$_{x}$CoO$_{3-\delta}$ ($0 < x
\leq 0.5$) \cite{itoh,ganguly,senaris,mahen,golovan}. Although the
system was studied for more than 50 years, not a few of its unique
magnetic and transport properties are still puzzling. The parent
compound LaCoO$_3$ is a non-magnetic insulator.  For the doping 
range $0.3 < x \leq 0.5$, the system is ferromagnetic (FM)
with the Curie temperature, $T_c$, in the range 
220-250~K~\cite{itoh,ganguly,senaris,mahen}.
\par
It should be pointed out that much more interest is being shown in
related FM perovskite-like oxides, mixed-valence manganites, of
the type La$_{1-x}$A$_x$MnO$_3$, where A is a divalent
alkaline-earth element like Ca, Sr, Ba \cite{coey,dagotto}. In
these compounds, the effect of so-called colossal
magnetoresistance (CMR) is found which offers applications in
advanced technology. It is usual, in this connection, when
considering properties of the cobaltites to compare them with
those of the manganites. The magnetoresistance (MR) in manganites,
defined as $\delta (H)=[R(0)-R(H)]/R(0)$, is found to be more than
90 \% at a field $H$ of about 60 kOe in the neighbourhood  of room
temperature. In spite of enormous efforts, a clear understanding
of CMR is not yet available, though some models describe
experimental data with an acceptable accuracy
\cite{coey,dagotto,beznos1}. Beside this, the requirement of high
magnetic fields hampers the possible application of the CMR
manganites.  In the Sr-doped cobaltites, the magnitude of the MR
is found to be rather low (only a few per cent). It is hoped that
elucidation of large difference in the MR between the manganites
and cobaltites could be helpful for understanding the nature of
CMR \cite{beznos1,beznos2}.
\par
The study presented in this article is inspired to a certain
extent by results of Ref.~\onlinecite{belev}, where transport
properties in a bulk ceramic sample and a film of
La$_{0.5}$Sr$_{0.5}$CoO$_{3-\delta}$ were measured by DC and
microwave (41 GHz) methods. It was found that the film presents a
highly inhomogeneous system of weakly connected FM grains. The
most important finding in Ref.~\onlinecite{belev} is the
following: the microwave conductivity of the film, which should be
related mainly to the conductivity within the poorly connected
grains, increases by an order of magnitude at the transition to
the FM state. This change is huge in comparison with the 10\%
decrease in the film DC resistivity found in the same temperature
range. The main conclusion of Ref.~\onlinecite{belev} is that the
microwave absorption  reflects more closely the temperature
variations in intragrain conductivity of inhomogeneous oxides than
the DC conductivity.
\par
It is known that absorption of electromagnetic waves in FM metals
is determined not only by the conductivity, but by the magnetic
permeability, $\mu$, of absorbing media as well. At a fairly high
frequency, the magnitude of $\mu$ is close to unity. The second
feature of the microwaves is their small penetration depth, which
is called the skin depth. In good conductors, the skin depth of
microwaves is a few micrometers. This is quite enough for 
thin films, but it often causes
 a problem in the study of bulk samples. The surface layer of
bulk samples can have differing properties from
the interior. This is especially true in the case of FM oxides
(including cobaltites), where a surface is often depleted in
oxygen \cite{belev}. It is advisable, in this connection, to use
low-frequency waves to study  bulk samples to ensure a fairly
large skin depth. But in this case the influence of permeability
on the absorption of these waves should be expected.
\par
It follows from the results of Ref.~\onlinecite{belev} that
high-frequency wave absorption might be much more sensitive to an
applied magnetic field than DC resistivity. The aim of this study
was to test this assumption. The subject of investigation was a
bulk sample of La$_{0.5}$Sr$_{0.5}$CoO$_{3-\delta}$.
Radio-frequency (RF) waves penetrating most of the sample volume
were used. We have  found a giant magnetoabsorption (about 38\%)
near the Curie temperature in a rather small magnetic field 2.1
kOe. This differs drastically from the measured MR values (the
$\delta (H)$ is about 0.26~\% in this field). The reasons
for the giant magnetoabsorption effect  are considered. 

\section{Experimental method}
The ceramic samples of La$_{0.5}$Sr$_{0.5}$CoO$_{3-\delta}$ are
all cut from the same pellet, prepared by a standard solid-state
reaction technique. The pellet was polycrystalline with grain size
in the range 40-70 $\mu$m. Some other details of the sample
preparation and characterization can be found in Ref.
\onlinecite{belev}. The measurements of the AC permeability were
done by an induction technique using a non-commercial magnetometer
in AC magnetic field $H_{\mathrm{ac}}=4$~Oe and frequency 10 kHz.
The DC resistance, as a function of temperature and magnetic field
$H$ (up to 20 kOe), was measured using a standard four-point probe
technique.
\par
The sample for RF measurements (with dimensions
21.6$\times$7$\times$4 mm$^3$) was placed in an induction coil
(9.7 mm in diameter and with a height approximately equal to
the longest side of the sample). This coil is a part of an LC tank
circuit. The magnitude of RF magnetic field in the induction coil
was not more than 0.1 Oe. The measurements of the quality factor,
$Q$, of the LC tank with and without the sample inside the coil
were taken. The measurements of the $Q$-factor depending on
temperature or magnetic field  were taken at fixed frequencies. In
this article, we present the results for the frequency $\nu
=1.33$~MHz. The available cryostat with electromagnet makes it
possible to measure the $Q$-factor in DC magnetic fields up to 2.5
kOe. The DC and RF fields were mutually perpendicular. The
temperature and magnetic-field dependences of the $Q$-factor were
recorded on heating after the sample had been cooled down in zero
field.
\par
Consider more exactly physical properties of the sample which can be
derived or, at least, judged by from the $Q$-factor measurements
of the RF LC tank employed. In the general case, the sample behavior
in an electromagnetic field can be described, using complex permittivity
$\epsilon_{\mathrm{c}} = \epsilon^{'}-\mathrm{i}\epsilon^{''} =
\epsilon^{'}\left[ 1-\mathrm{i}\tan (\delta_{\epsilon})\right]$
(where $\epsilon^{'}=\epsilon_{r}\epsilon_{0}$,
$\tan (\delta_{\epsilon}) = \sigma_{\mathrm{rf}}/(\epsilon^{'} \omega)$
is the dielectric loss tangent,
$\sigma_{\mathrm{rf}}$ is the RF conductivity of the sample) and complex
magnetic permeability $\mu_{\mathrm{c}}=\mu^{'} - \mathrm{i}\mu^{''} =
\mu^{'}\left[ 1-\mathrm{i}\tan(\delta_{\mu})\right]$, where
$\mu^{'}=\mu_{\mathrm{r}}\mu_{0}$ and $\tan(\delta_{\mu})=
\mu^{''}/\mu^{'}$.
For a plane electromagnetic wave, the complex resistance of the sample
is given by relation
\begin{equation}
Z_{\mathrm{c}}= \sqrt{\frac{\mu_{\mathrm{c}}}{\epsilon_{\mathrm{c}}}} =
Z_0
\sqrt{\frac{\mu_{\mathrm{r}}}{\epsilon_{\mathrm{r}}}
\left ( \frac{1-\mathrm{i}\mu^{''}/\mu^{'}}
{1-\mathrm{i}\sigma_{\mathrm{rf}}/\epsilon^{'} \omega} \right )}
=R_{\mathrm{c}}+\mathrm{i}X_{\mathrm{c}},
\end{equation}
where $Z_0=120$~$\Omega$,
$R_{\mathrm{c}}$ defines the energy loss in the sample,
$X_{\mathrm{c}}$ is the sample reactance.

\par
Once the sample with the wave resistance $Z_{\mathrm{c}}$ is inserted
into the coil of the RF circuit, arranged from resistance, inductance
and capacitance elements $R_0$, $L_0$ and $C_0$, a complex resistance
is added to the circuit, given by
\begin{equation}
Z_{\mathrm{sp}}=kZ_{\mathrm{c}}=R_{\mathrm{sp}}+\mathrm{i}X_{\mathrm{sp}},
\end{equation}
where $k$ is a geometrical factor determined by the sample and coil
dimensions, which was about 0.38 in this study. The
$Q$-factor of the circuit without the sample is expressed as
$Q_0 = \omega L_{0}/R_0$.
It is often more convenient to use the  damping factor (or decrement)
$d=Q^{-1}$. The decrement of the circuit without the sample is, therefore,
$d_0=R_{0}/\omega L_{0}$. On inserting the sample into the coil the circuit
decrement changes to
$
d=(R_{0}+kR_{\mathrm{c}})/[\omega(L_{0}+kL_{\mathrm{c}})].
$
In the case that the sample causes merely small changes in the
circuit inductance ($kL_{\mathrm{c}}\ll L_{0}$) and the circuit
quality without the sample is fairly high ($\omega L_{0}/R_{0} \gg
1$), the expression for the difference $d-d_0=P_{\mathrm{A}}$ can
be written as
\begin{equation}
P_{\mathrm{A}}=k\frac{R_{\mathrm{c}}}{\omega L_{0}} =k\frac{R_{\mathrm{c}}}
{R_{0}Q_{0}}.
\label{eq}
\end{equation}
It is seen from it that $P_{\mathrm{A}}$ is determined by the loss
in the sample. For the further analysis of the situation it is
useful to consider the following approximations: $\tan
(\delta_{\epsilon})\gg 1$ and $\tan(\delta_{\mu})\ll 1$. This
enables us to write the Eq. \ref{eq} in the form
\begin{equation}
P_{\mathrm{A}}=k_0 \left (1+\frac{1}{2}\frac{\mu^{''}}{\mu^{'}}\right )
R_s
\label{p2}
\end{equation}
where $k_{0} =k/(\omega L_{0})$,
$R_s=(\omega\mu^{'}/2\sigma_{\mathrm{rf}})^{1/2}=
(\omega\mu^{'}\rho_{\mathrm{rf}}/2)^{1/2}$ is the surface
resistance, $\rho_{\mathrm{rf}}$ is the RF resistivity. The Eq.
\ref{p2} present a simplified way to take into account the
influence of the conductivity and permeability on the RF loss.

\section{Results and discussion}
\begin{figure}
\centerline{\epsfig{file=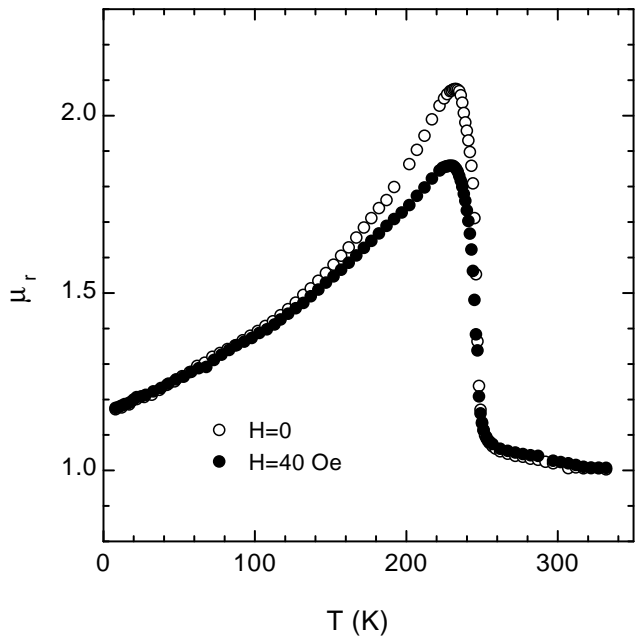,width=9cm}}
\caption{Temperature dependences of the real part of the relative
magnetic permeability, $\mu_{\mathrm{r}}=\mu^{'}/\mu_{0}$, of the sample
La$_{0.5}$Sr$_{0.5}$CoO$_3$ in AC magnetic field
$H_{\mathrm{ac}}=4$~Oe and frequency 10 kHz. The dependences have been
recorded in zero field (open circles) and in DC field 40 Oe (solid circles)
with temperature increasing after the sample was cooled in zero
field.}
\label{mut}
\end{figure}

The temperature dependence of the real part, $\mu_{\mathrm{r}}$,
of the relative AC permeability of the sample studied is shown in
Fig.~\ref{mut}. The value of $T_{\mathrm{c}}\approx 250$~K is
found if $T_{\mathrm{c}}$ is defined as the temperature of the
inflection point in the $\mu_{\mathrm{r}}(T)$ curve. In general,
the behavior of $\mu_{\mathrm{r}}(T)$ agrees well with the known
studies of the AC susceptibilty in ceramic
La$_{0.5}$Sr$_{0.5}$CoO$_{3-\delta}$ \cite{ganguly,mukh,anil}.
When going from high to lower temperature, the relative AC
permeability increases slightly between 320 K and 250~K, then it
increases sharply at $T =T_{\mathrm{c}}\approx 250$~K and reaches
its maximum at a temperature $T_{\mathrm{max}}= 232$~K. A
considerable decrease in $\mu_{\mathrm{r}}(T)$ (or AC
susceptibility) below $T_{\mathrm{max}}$ (Fig.~\ref{mut}) is a
well-known feature of cobaltites. It is attributed to the
cluster-glass behavior \cite{ganguly} or to an increase in the
coercivity with decreasing temperature  \cite{anil}.
Figure~\ref{mut} demonstrates a high sensitivity of the
$\mu_{\mathrm{r}}(T)$ to an external magnetic field: even a fairly
low field (40 Oe) causes a considerable decrease in
$\mu_{\mathrm{r}}(T)$. The effect is maximal at
$T=T_{\mathrm{max}}$. When going to either side from
$T_{\mathrm{max}}$, the influence of the magnetic field becomes
weaker. In particular, it is negligible for $T\leq 100$~K.

\begin{figure}[h] 
\vspace{-15pt}
\centerline{\epsfig{file=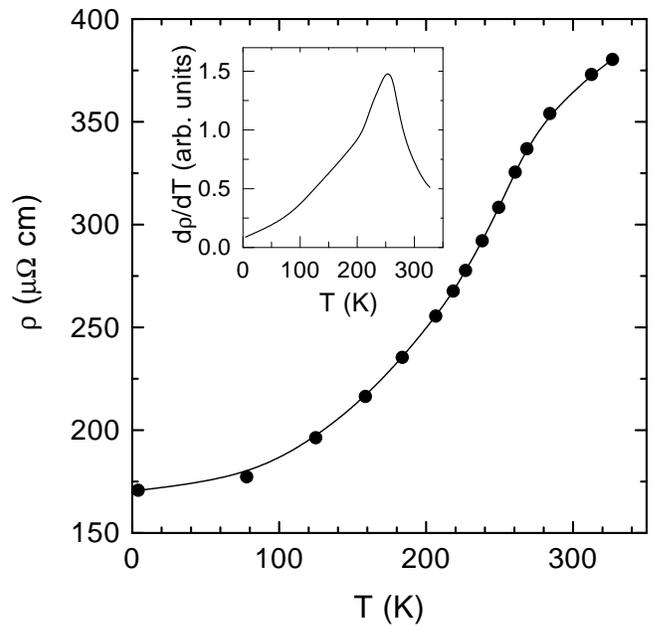,width=9.5cm}}
\vspace{-15pt}
\caption{Temperature dependence of the DC resistivity and its
derivative $\mathrm{d}\rho/\mathrm{d}T$ (insert) of the
sample La$_{0.5}$Sr$_{0.5}$CoO$_3$.}
\label{rt}
\end{figure}
\begin{figure}
\centerline{\epsfig{file=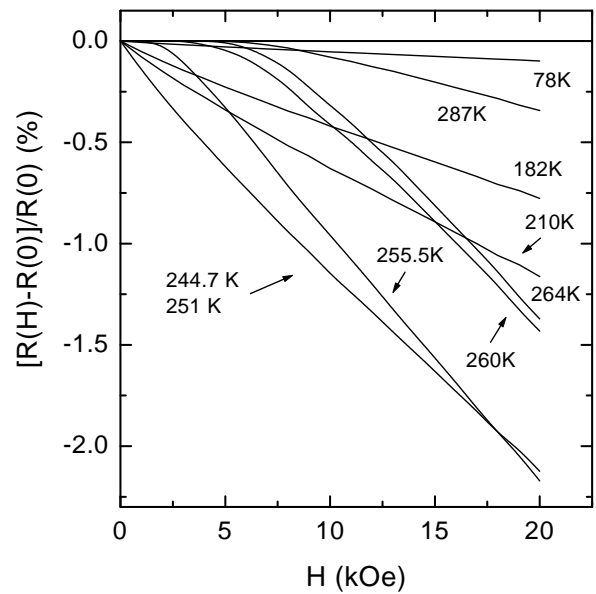,width=9.2cm}}
\caption{Magnetoresistance of the sample studied for different
temperatures.}
\label{rh}
\end{figure}
\par
\begin{figure}[h]
\vspace{-12pt}
\centerline{\epsfig{file=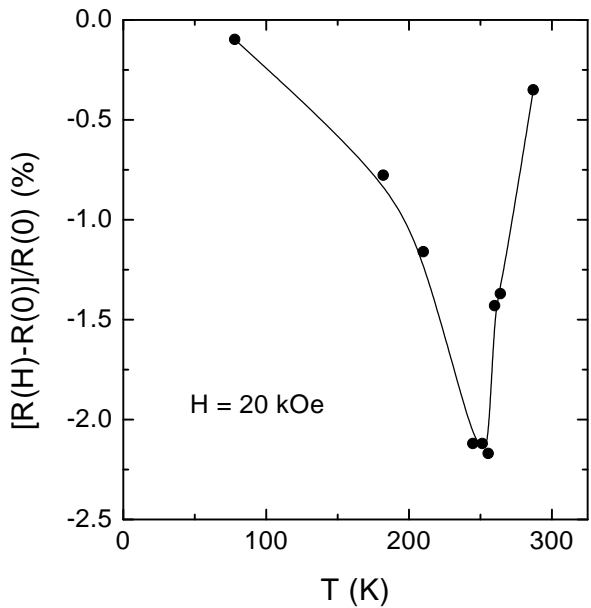,width=9.5cm}}
\vspace{-15pt}
\caption{Temperature dependence of the magnetoresistance at $H=20$~kOe.
The solid line presents a B-spline fitting.}
\label{mrt}
\end{figure}

The temperature dependence of the DC resistivity, $\rho (T)$, is
shown in Fig.~\ref{rt}. The $\rho$-values were calculated taking
into account the porosity of the sample \cite{belev}. The $\rho
(T)$ behavior is metallic ($\mathrm{d}\rho/\mathrm{d}T > 0$) over
the whole temperature range investigated, below and above
$T_{\mathrm{c}}$.  This behavior (and the resistivity values)
agree with the known data on ceramic La$_{0.5}$Sr$_{0.5}$CoO$_3$
\cite{ganguly}. The $\rho (T)$ dependence exhibits a change of
slope at $T_{\mathrm{c}}$ (see insert in Fig.~\ref{rt}) due to a
contribution from the electron scattering on the spin disorder (in
addition to the usual contributions from crystal lattice defects
and electron-phonon scattering) \cite{vons}. This ``magnetic''
contribution to the resistivity, $\Delta \rho_{m}$, depends on the
magnetization. The external magnetic field enhances the spin order
(that is, the magnetization), which leads to a decrease in the
resistivity, that is to negative MR. The magnetic-field
dependences of resistance of the sample studied are shown in
Fig.~\ref{rh}. It is seen that the MR depends strongly on
temperature. The temperature dependence of the MR at $H=20$~kOe is
presented in Fig.~\ref{mrt}. This behavior is quite common for
FM oxides (like manganites and cobaltites). The MR is maximal at
$T\approx T_{\mathrm{c}}$. It goes down rather steeply for
temperature deviating to either side from $T_{\mathrm{c}}$. This
temperature behavior of the MR is expected for FM oxides of fairly
good crystal perfection or for polycrystalline samples with rather
strong connectivity between the crystallites. Indeed, the MR
amplitude is determined by the ability of an external magnetic
field to increase the magnetization. It is obvious that at low
temperatures ($T\ll T_{\mathrm{c}}$), when nearly all spins are
already aligned by the exchange interaction, this ability is
minimal. For increasing temperature and, especially, at
temperature close to $T_{\mathrm{c}}$, the magnetic order becomes
weaker (the magnetization goes down) due to thermal fluctuations.
In this case the possibility to strengthen the magnetic order with
an external magnetic field increases profoundly. This is the
reason for maximal MR magnitude near $T_{\mathrm{c}}$. Above
$T_{\mathrm{c}}$, the spin arrangement becomes essentially random,
the magnetization is zero, and, therefore, the MR is close to zero
as well.
\begin{figure}
\vspace{-17pt}
\centerline{\epsfig{file=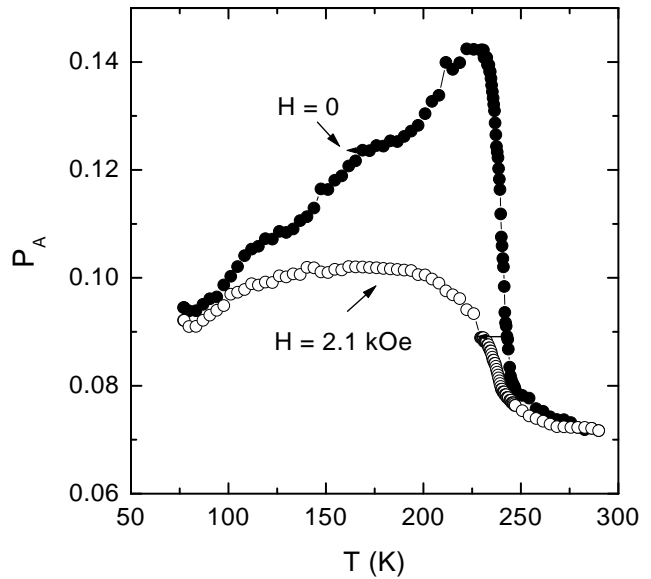,width=9cm}}
\vspace{-15pt}
\caption{Temperature dependences of RF absorption ($\nu =1.33$~MHz) in
the sample studied  in zero field and in the field $H=2.1$~kOe.
$P_{\mathrm{A}}= d-d_0$, where $d$ and $d_0$
are the decrements of the used LC circuit with the sample in an inductance
coil and without the sample in it, respectively.}
\vspace{-5pt}
\label{at}
\end{figure}

\par
Consider now the behavior of the RF loss $P_{\mathrm{A}}$ in the
sample studied. We begin with the temperature behavior
$P_{\mathrm{A}}(T)$ in zero field (Fig.~\ref{at}). In doing so
we will refer to Eq.~ \ref{p2}. It is worthwhile to mention here
that certain mutual correlations in the temperature behavior of
$\mu^{'}$ and $\mu^{''}$ can be expected. It follows immediately
from the Kramers-Kronig relations \cite{white}. These are
helpful in consideration of a remarkable correlation between the
temperature behavior of the loss and the permeability
$\mu_{\mathrm{r}}$ found in this study. The correlation is evident
at once on comparing Figs. \ref{mut} and \ref{at}. It is clear
that far enough above  $T_{\mathrm{c}}$ the relations,
$\mu_{\mathrm{r}}=1$ and $\mu^{''}=0$, are true. In this case,
according to Eq.~\ref{p2}, the loss depends only on the
conductivity. In FM oxides, like mixed-valence manganites and
cobaltites, the FM fluctuations (or developing of small FM regions
in the paramagnetic matrix) can be evident, however, fairly far
above $T_{\mathrm{c}}$ for extrinsic and intrinsic sources of
magnetic inhomogeneity \cite{itoh,senaris,mahen,dagotto,belev}. In
that event, as the $T_{\mathrm{c}}$ is approached from above, the 
slight increase in the susceptibility or permeability can be 
observed even in the range far from $T_{\mathrm{c}}$. This will 
cause a corresponding increase in the loss. The drastic increase in
$P_{\mathrm{A}}$ occurs, however, only quite near $T_{\mathrm{c}}$
due the sharp increase in $\mu^{'}$ at transition of the most of
the sample volume to the FM state (Figs.~\ref{mut} and \ref{at}). 
Notice that the conductivity
increases with temperature decreasing in the whole temperature
range studied (Fig.~\ref{rt}) and this (according to the 
Eq.~\ref{p2}) favors a decrease in the loss. Contribution of the
permeability to temperature variations of the loss is, however,
far more than that of the conductivity, so, as a result, the loss
increases at the transition to the FM state. For further decrease in
temperature below $T_{\mathrm{c}}$ the loss comes to some maximal
value at $T\approx 230$~K and, after that, goes down
(Fig.~\ref{at}). All this correlates well with the
$\mu_{\mathrm{r}}(T)$ behavior (Fig.~\ref{mut}). Even the
temperatures of the maxima for $P_{\mathrm{A}}$ and
$\mu_{\mathrm{r}}(T)$ curves are both at about 230~K.
\begin{figure}
\vspace{-17pt}
\centerline{\epsfig{file=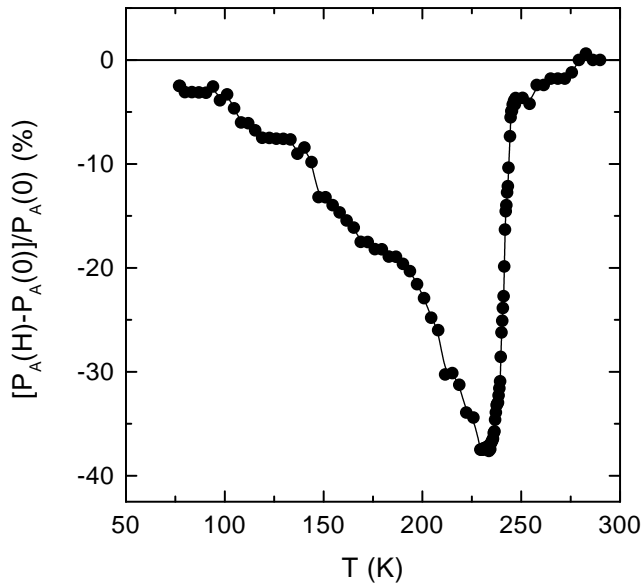,width=9cm}}
\caption{Temperature dependence of the RF magnetoabsorption in the
sample studied in field $H=2.1$~kOe.}
\label{mat}
\end{figure}
\par
It is seen that the RF loss is far more sensitive to strengthening
of the spin order at the transition to the FM state than the DC
conductivity. Really, $\rho (T)$ shows only a change in the slope
at $T_{\mathrm{c}}$ that means the steeper decrease in resistivity
with temperature decreasing below  $T<T_{\mathrm{c}}$
(Fig.~\ref{rt}). In contrast to this, the RF loss increases
sharply (nearly twice) in the vicinity of $T_{\mathrm{c}}$ in zero
field (Fig.~\ref{at}). This indicates that it is precisely the
changes in the permeability that cause the large variations in the
loss near $T_{\mathrm{c}}$.
\par
\begin{figure}[h]
\vspace{-12pt}
\centerline{\epsfig{file=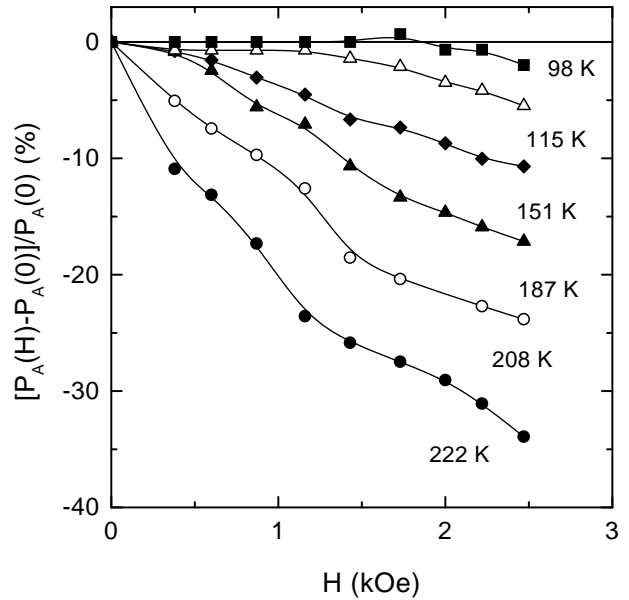,width=9cm}}
\vspace{-10pt}
\caption{Magnetic-field dependences of the RF magnetoabsorption in the
sample studied for different temperatures. The solid lines are guides
to the eye.}
\label{mah}
\end{figure}

From measurements of the loss, $P_{\mathrm{A}}$, at different
magnetic fields we have obtained the values of the
magnetoabsorption (MA), which has been defined as
$[P_{\mathrm{A}}(H)-P_{\mathrm{A}}(0)]/P_{\mathrm{A}}(0)= \Delta
P_{\mathrm{A}}(H)/P_{\mathrm{A}}(0)$. Using the above-mentioned
approximation, $\tan(\delta_{\mu})= \mu^{''}/\mu^{'}\ll 1$, and
taking into account Eq.~\ref{p2}, a simplified expression for MA
can be written as
\begin{equation}
\Delta P_{\mathrm{A}}(H)/P_{\mathrm{A}}(0)\approx
\frac{R_{s}(H)}{R_{s}(0)}-1= \sqrt{\frac{\mu^{'}(H)\rho_{\mathrm{rf}}(H)}
{\mu^{'}(0)\rho_{\mathrm{rf}}(0)}}-1.
\label{ph}
\end{equation}
This shows immediately the correlation between the
$P_{\mathrm{A}}(T)$ and $\mu^{'}(T)$. It is clear from
Eqs.~\ref{p2} and \ref{ph} that owing to decrease in
$\rho_{\mathrm{rf}}$ and $\mu^{'}$ in an applied magnetic field,
the MA should be negative. This is consistent with the results
found. We actually found a high sensitivity of the RF loss to
applied magnetic field (Figs.~\ref{at}, \ref{mat},
\ref{mah}). It is remarkable that the strong magnetoabsorption
effect shows itself in rather low magnetic fields. The effect is
really giant when compared with the DC magnetoresistance. Not far
from $T_{\mathrm{c}}$, at $T_{\mathrm{max}}\approx 230$~K, the MA
is about 38\% in field $H=2.1$~kOe (Fig.~\ref{mat}). By
contrast, the MR is about 0.26 \% near $T_{\mathrm{c}}$ in
the above-mentioned field (Fig.~\ref{rh}) and it has increased
no more than to about 2.15~\% at field $H=20$~kOe
(Figs.~\ref{rh} and \ref{mrt}).
\par
The reason why the magnetoabsorption effect is far stronger than
the MR one appears rather obvious. As indicated above, the MR in
the vicinity and below $T_{\mathrm{c}}$ is determined by the
enhancing of the spin order (magnetization) in an applied magnetic
field. This leads to the conductivity increasing (that is to the
negative MR). Unlike the DC magnetoresistance, the
magnetoabsorption is determined not only by influence of the
magnetic field on the conductivity, but, to a much greater extent,
by changes in the permeability in the magnetic field.
The remarkable correlation between behaviors of the
$P_{\mathrm{A}}(T)$ and the $\mu_{\mathrm{r}}(T)$ was mentioned
above. No less noteworthy correlation can be found between the
behaviors of $P_{\mathrm{A}}$ and the $\mu_{\mathrm{r}}$ in an
applied magnetic field (compare Figs.~\ref{mut} and
\ref{mat}).

\par
It follows that the giant magnetoabsorption effect is qualitatively 
explicable taking into account a decrease in the DC permeability. The 
RF resistivity, $\rho_{\mathrm{rf}}$,
decreases in magnetic field as well and, therefore, gives also
some contribution to this effect. It could be expected
\cite{belev}, that RF resistivity is more sensitive to magnetic
field than DC resistivity. We cannot, however, separate a
contribution of RF resistivity to the magnetoabsorption. In any
case, however, the above-described results testify that this
contribution is negligible as compared to that of the
permeability.
\par
No direct measurements of $\mu^{''}$ in the
La$_{0.5}$Sr$_{0.5}$CoO$_{3-\delta}$ or other Sr-doped cobaltites
have been made, to our knowledge. It is known that the value of
$\mu^{'}$ decreases with increasing frequency \cite{vons}, but it can be
large for fairly low frequencies. This tendency, general for FM metals, is
found to hold for related perovskite-like oxides as well. For example, for
polycrystalline manganite La$_{0.3}$Sr$_{0.7}$MnO$_{3}$ the magnitude of
$\mu^{'}$ is fairly close to unity at $\nu=100$~MHz, but in the 1-MHz-range
it can be significantly larger than unity \cite{wang1}. In this case the
relation, $\mu^{''}/\mu^{'}\ll 1$, holds.
\par
In conclusion, we found the giant negative RF magnetoabsorption
effect in bulk sample of La$_{0.5}$Sr$_{0.5}$CoO$_{3-\delta}$
under low DC magnetic fields. The effect should be attributed
primarily to the changes in the magnetic permeability in applied
magnetic field. Such low-field magnetoabsorption may be of
potential application in RF devices controlled by magnetic field
\cite{knobel}. It is clear that effective temperature range for 
application of any material with giant magnetoabsorption is determined by
its $T_{c}$ value. For the range near or above room temperature, the 
manganite La$_{1-x}$Sr$_{x}$MnO$_{3}$ with $0.25 < x < 0.35$ 
is suggested.



\newpage

\begin{thebibliography}{00}

\bibitem{itoh}M. Itoh, I. Natori, S. Kubota, and K. Motoya,
J. Phys. Soc. Jap. {\bf 63,} 1486 (1994).

\bibitem{ganguly}P. Ganguly, P. S. Anil Kumar, P. N. Santosh and
I. S. Mulla, J. Phys.: Condens Matter {\bf 6,} 533 (1994).

\bibitem{senaris}M. A. Se\'{n}aris-Rodriguez and J. B. Goodenough,
J. Sol. State Chem. {\bf 118,} 323 (1995).

\bibitem{mahen}R. Mahendiran and A. K. Raychaudhuri, Phys. Rev. B
{\bf 54,} 16044 (1996).

\bibitem{golovan}V. Golovanov, L. Mihaly, and A. R. Moodenbaugh,
Phys. Rev. B {\bf 53,} 8207 (1996).

\bibitem{coey}J. M. D. Coey, M. Viret, and S. von Molnar, Adv. Phys. {\bf 48,}
167 (1999).

\bibitem{dagotto}E. Dagotto, T. Hotta, and A. Moreo, Phys. Rep. {\bf 344,}
1 (2001).

\bibitem{beznos1}A. B. Beznosov, B. I. Belevtsev, E. L. Fertman,
V. A. Desnenko, D. G. Naugle, K. D. D. Rathnayaka and A. Parasiris,
Fiz. Nizk. Temp., {\bf 28,} 774 (2002) [Low Temp. Phys., {\bf 28,}
556 (2002)].

\bibitem{beznos2}A. B. Beznosov, P. P. Pal-Val, B. I. Belevtsev,
V. B. Krasovitsky, E. L. Fertman, L. N. Pal-Val, I. N. Chukanova and
T. G. Deyneka, Izv. AN, Ser. Fiz., {\bf 66,} 758 (2002).

\bibitem{belev} B. I. Belevtsev, N. T. Cherpak, I. N. Chukanova, A. I. Gubin,
V. B. Krasovitsky, and A. A. Lavrinovich, J. Phys.: Condens. Matter
{\bf 14,} 2591 (2002).

\bibitem{mukh} S. Mukherjee, R. Ranganathan, P. S. Anilkumar, and
P. A. Joy, Phys. Rev. B {\bf 54,} 9267 (1996).

\bibitem{anil}P. S. Anil Kumar, P. A. Joy, and S. K. Date,
J. Phys.: Condens. Matter {\bf 1,} L487 (1998).

\bibitem{vons}S. V. Vonsovsky, {\it Magnetism,} (Nauka, Moscow, 1971).

\bibitem{white}R. M. White, {\it Quantum theory of magnetism,}
(Springer-Verlag, New York, 1983).

\bibitem{wang1}Jinhui Wang, Gang Ni, Wenli Gao, Benxi Gu, Xiabin Chen, and
Youwei Du, phys. stat. sol. (a) {\bf 183,} 421 (2001).

\bibitem{knobel}M. Knobel and K. R. Pirota, J. Magn. Magn. Mater.
{\bf 242-245,} 33 (2002).

\end{thebibliography}

\end{document}